# A hydrophobic-interaction-based mechanism trigger docking between the SARS-CoV-2 spike and angiotensin-converting enzyme 2

Jiacheng Li[a,1], Xiaoliang Ma[a,1], Shuai Guo[a,1], Chengyu Hou[b,1], Liping Shi[a], Hongchi Zhang[a],

Bing Zheng[c], Chencheng Liao[b], Lin Yang[a,d,*], Lin Ye[d], Xiaodong He[a,e]

[a] *National Key Laboratory of Science and Technology on Advanced Composites in Special Environments, Center for Composite Materials and Structures, Harbin Institute of Technology, Harbin 150080, China*

[b] *School of Electronics and Information Engineering, Harbin Institute of Technology, Harbin 150080, China*

[c] *Key Laboratory of Functional Inorganic Material Chemistry (Ministry of Education) and School of Chemistry and Materials Science, Heilongjiang University, Harbin 150001, P. R. China.*

[d] *School of Aerospace, Mechanical and Mechatronic Engineering, The University of Sydney, NSW 2006, Australia*

[e] *Shenzhen STRONG Advanced Materials Research Institute Co., Ltd, Shenzhen 518035, P. R. China.*

## Abstract

A recent experimental study found that the binding affinity between the cellular receptor human angiotensin converting enzyme 2 (ACE2) and receptor-binding domain (RBD) in spike (S) protein of novel severe acute respiratory syndrome coronavirus 2 (SARS-CoV-2) is more than 10-fold higher than that of the original severe acute respiratory syndrome coronavirus (SARS-CoV). However, main-chain structures of the SARS-CoV-2 RBD are almost the same with that of the SARS-CoV RBD. Understanding physical mechanism responsible for the outstanding affinity between the SARS-CoV-2 S and ACE2 is the "urgent challenge" for developing blockers, vaccines and therapeutic antibodies against the coronavirus disease 2019 (COVID-19) pandemic. Considering the mechanisms of hydrophobic interaction, hydration shell, surface tension, and the shielding effect of water molecules, this study reveals a hydrophobic-interaction-based mechanism by means of which SARS-CoV-2 S and ACE2 bind together in an aqueous environment. The hydrophobic interaction between the SARS-CoV-2 S and ACE2 protein is found to be significantly greater than that between SARS-CoV S and ACE2. At the docking site, the hydrophobic portions of the hydrophilic side chains of SARS-CoV-2 S are found to be involved in the hydrophobic interaction between SARS-CoV-2 S and ACE2. We propose a method to design live attenuated viruses by mutating several key amino acid residues of the spike protein to decrease the hydrophobic surface areas at the docking site. Mutation of a small amount of residues can greatly reduce the hydrophobic binding of the coronavirus to the receptor, which may be significant reduce infectivity and transmissibility of the virus.

*Corresponding author. E-mail address: linyang@hit.edu.cn (Lin Yang) [1]These authors contributed equally to this work.

**Introduction**

The novel severe acute respiratory syndrome coronavirus 2 (SARS-CoV-2) has emerged as a human pathogen, causing fever, severe respiratory diseases, pneumonia, and systemic inflammatory response syndrome, leading to a worldwide sustained pandemic. Both SARS-CoV-2 and the original severe acute respiratory syndrome coronavirus (SARS-CoV) enter human cells by protein-protein docking to human angiotensin converting enzyme 2 (ACE2) on the host cell membrane via CoV spike (S) glycoproteins. A recent experimental study found that the binding affinity between ACE2 and the receptor-binding domain (RBD) of the S protein of SARS-CoV-2 is more than 10-fold higher than that of SARS-CoV, which may contribute to the higher infectivity and transmissibility of SARS-CoV-2 compared to SARS-CoV (*1-3*).

Molecular structures of the S protein of SARS-CoV-2 have been observed at high resolution by using cryo-electron microscopy (cryo-EM) (*4-6*). The complex structures of ACE2 bound to the SARS-CoV-2 S have also been experimentally determined (*7-10*). Surprisingly, all these experiments showed that the backbone structures of the RBD of SARS-CoV-2 S are almost same as that of SARS-CoV S (see Fig. 1a) (*7, 11*). A molecular dynamic (MD) study has showed that the binding energy of SARS-CoV-2 S to ACE2 is almost same as that of SARS-CoV S to ACE2 (*12*). Another MD simulation study showed that the interaction ability between SARS-CoV-2 RBD and ACE2 decreased by 35.6% compared with the interaction ability of SARS-CoV RBD and ACE2, attributed to the lack of several hydrogen bonds between SARS-CoV-2 RBD and ACE2, the molecular binding free energy is therefore significantly reduced (*13*). Therefore, physical mechanisms responsible for the strong binding affinity between SARS-CoV-2 RBD and ACE2 haven't be disclosed by the binding energy calculations. The reason why the affinity of SARS-CoV-2 RBD and ACE2 far exceeds that of SARS-CoV RBD and

ACE2 may be caused by a long-range adhesion mechanism between the ligand and receptor.

Specific binding of SARS-CoV-2 S and ACE2 forms the joint structure between the coronavirus and the host cell that enable the coronavirus enter the host cell (*14*). The chief characteristic of proteins that allows their diverse set of functions is their ability to dock with other proteins specifically and tightly. Protein-protein docking is therefore considered one of the miracles of nature, in that almost all biological existence, functionalization, diversity, and evolution rely on it as the most important mechanism, principle, and motivation. At present the underlying physical mechanisms responsible for the specific docking of SARS-CoV-2 RBD and ACE2 are not fully understood (*15*), which hinders the development of anti-coronavirus drugs and therapies. Surprisingly, in natural intracellular environment and extracellular medium, protein-protein docking are usually the contacts of high specificity established between two or more specific protein molecules, and erroneous protein-protein docking rarely occurs (*16*). The classic problem of protein-protein docking is the question of how a protein find its partner in its natural environment (*17*).

Protein-protein docking is mainly guided by a variety of physical forces as follows: (i) hydrophobic effect, (ii) electrostatic forces, (iii) van der Waals forces, (iv) hydrogen bonding, (v) ionic bonding, (vi) entropy. Among them, the hydrogen bonding and hydrophobic effect are normally thought to play a decisive role (*15, 18*). In extracellular medium, hydrogen-bond competing is always present with water. Because bulk water interferes with reversible biological processes and enthalpy-entropy compensation occurs during hydrogen-bond formation, the mechanisms and the extent to which hydrogen bonds contribute to protein-protein docking are not well understood. In particular, whether hydrogen bonds formation regulate protein-protein docking remains a long-standing problem with poorly defined mechanisms (*15, 19-23*). It is worth noting that hydrogen bonds formation is not a long-range physical force. Considering that only several hydrogen bonds between SARS-CoV-2 RBD and ACE2 can be identified in the complex (*7-10, 13*), the docking between the coronavirus and the host cell may be not dominated by hydrogen bond pairing between them.

Water molecules have a very strong polarity (*24*). The interaction of protein surface with the surrounding water is often referred to as protein hydration layer (also sometimes called hydration shell) and is fundamental to structural stability of protein, because non-aqueous solvents in general denature proteins (*25*). The hydration layer around a protein has been found to have dynamics distinct from the bulk water to a distance of 1 nm and water molecules slow down greatly when they encounter a protein(*26*). Thus, hydrophilic side chains of proteins are normally hydrogen bonded with surrounding water molecules in aqueous environments, thereby preventing the surface hydrophilic side-chains of proteins from randomly hydrogen bonding together (*26, 27*) (*24*). This is the reason why proteins usually do not aggregate and crystallize in unsaturated aqueous solutions(*28*).

The region of the protein responsible for binding another molecule is known as the docking site (also sometimes called binding site) and is often a depression on the molecular surface. Before the docking, external hydrophilic side-chains of SARS-CoV-2 S and ACE2 must hydrogen bonded with water molecules in extracellular medium, so it is difficult to explain how the hydrophilic side-chains at the docking site can get rid of their hydrogen-bonded water molecules, and then interact with each other during the docking process (*7-10*).

The key to SARS-CoV-2 infection is that the S protein can specifically bind to the ACE2 in a strong affinity manner. This binding ability is mediated by the tertiary structure of the protein, which defines the docking site, and by the chemical properties of the surrounding amino acids' side chains (*29*). The hydrophobicity of the protein surface is the main factor that stabilizes the protein-protein binding, thus hydrophobic interaction among proteins may play an important role in determining the protein-protein binding affinity (*18, 30, 31*)。

**Results**

Although there are many hydrophilic side chains on the surface of SARS-CoV-2 RBD, the surface of SARS-CoV-2 RBD is not completely hydrophilic. Hydrophobic side-chains of many species of residues, such as glycine (Gly), alanine (Ala), valine (Val), leucine (Leu), isoleucine (Ile), proline (Pro), phenylalanine (Phe), methionine (Met), and tryptophan (Trp) are found on the surface of the RBD, as shown in Fig. 1b (*4-10*). It is worth noting that hydrophilic side-chains are not completely hydrophilic. The hydrophilicity of hydrophilic side-chains is normally expressed by C=O or N-H2 groups at their ends, and the other portions of hydrophilic side-chains are hydrophobic, because the molecular structures of these portions are basically alkyl and benzene ring structures, as shown in Fig. 2. It means that there are a large number of water molecules surround the hydrophobic surface areas of the RBD rather than hydrogen bonded with the RBD. The characteristic of these water molecules surrounding the hydrophobic surface areas is that their hydrogen bonding network is more ordered than free liquid water molecules, that is, their entropy is lower.

We simulated the hydration layer of SARS-CoV-2 RBD by using the molecular dynamic method. The simulation results show that only about 30.6% of the water molecules in the innermost hydration layer surrounding the RBD hydrogen bonded with the RBD (see Fig. 3), due to exposure of many hydrophobic surface areas on the RBD (*4-10, 18, 30*). Many hydrophobic areas on the surface of the RBD are found to be connected with each other, which indicates that surface tension affect the surface properties of docking site of the RBD (see Fig. 1b) (*32*). Hydration layer of an ACE2 is also obtained by using MD simulation, showing that only about 21.3% of the water molecules in the innermost hydration layer surrounding the ACE2 hydrogen bonded with the ACE2 (see Fig. 4). The existence of hydrophobic surface in large areas of both SARS-CoV-2 RBD and ACE2 indicates that a strong hydrophobic interaction may occur between them.

To illustrate hydrophobic binding effect in the complex of SARS-CoV-2 RBD and ACE2, we mark the hydrophobic surface areas of the two proteins at the docking site based on the experimentally determined structure as shown in Fig. 5 (*9*). By analyzing the details of the interface between the RBD and ACE2 of the complex, it can be easily found that the docking causes the hydrophobic surface areas of the two protein contact and collapse together at the docking site. The hydrophobic interaction surface areas of the RBD to ACE2 accounts for 76% of the total contact area of the RBD (see Fig. 5). The degree of hydrophobic paring is very high and the hydrophobic interaction most likely play an important role in the protein-protein docking (*18, 30*). The hydrophobic portions of hydrophilic side-chain obviously participate in the hydrophobic interaction between the

RBD and ACE2 at the docking site (see Fig. 5). It is most likely that the hydrophobic interaction at the docking site enable the hydrophilic side-chains to get rid of their original hydrogen-bonded water molecules, so that the hydrophilic side-chains can participate in the hydrophobic interaction via their hydrophobic portions, namely enthalpy-entropy compensation occurs during the docking (*15*) (*19-23*).

Comparing the experimentally determined molecular structure of the complex of SARS-CoV RBD and ACE2, the corresponding hydrophobic surface areas of the two proteins at the docking site that involved in the hydrophobic binding interaction are marked in Fig. 6. By analyzing the docking site, it can be found that the hydrophobic surface areas involved in hydrophobic effect of the docking of SARS-CoV RBD and ACE2 is significantly smaller than that of the docking of SARS-CoV-2 RBD and ACE2 (see Fig. 6). Because many hydrophobic surface areas on ACE2 face to face with hydrophilic groups of SARS-CoV-2 RBD at the docking site. This means that the hydrophobic interaction between the SARS-CoV RBD and ACE2 is significantly less than that of between SARS-CoV-2 RBD and ACE2, due to the relatively poor hydrophobic pairing between SARS-CoV RBD and ACE2. This explains why the SARS-CoV-2 S exhibits a much higher affinity to the ACE2 protein than the SARS-CoV. We calculate the size of hydrophobic surface areas of SARS-CoV-2 RBD and SARS-CoV RBD at binding site that participate in the hydrophobic binding interaction with ACE2. The hydrophobic surface area of SARS-CoV-2 RBD (about 867.4Å2) that involved in the hydrophobic interaction docking with ACE2 is about 2.03 times of that (about 427.3Å2) of the SARS-CoV RBD (see Fig. 5,6).

We simulate the hydration layers of the complex of ACE2 bound to SARS-CoV-2 RBD and the complex of ACE2 bound to SARS-CoV RBD by using MD method, respectively. Two cross-sectional views of the two hydration shells at the docking sites are shown in FIG. 7. From the cross-sectional view of the complex of ACE2 and SARS-CoV-2 RBD, we can see that the docking causes the hydration shells of the RBD and ACE2 to be integrated. This means that the docking cause many ordered water molecules in the original hydration shells of the RBD and ACE2 at the docking site transformed into free water molecules, driven by an increase in entropy. At the docking site of the RBD and

ACE2, the side-chains of those hydrophilic residues have lost their original hydrogen-bonded water molecules and formed new hydrogen bonding, electrostatic interaction and hydrophilic interaction with each other, which results in the contact area of hydrophobic interaction at the docking site increasing. By comparing the hydrophobic surface areas of ACE2 at the docking site before and after docking with SARS-CoV-2 RBD, we found that the docking causes some disconnected hydrophobic surface areas at the docking site to be connected. Above all, it can be considered that the docking of SARS-CoV-2 RBD and ACE2 is mainly regulated by the hydrophobic effect at the binding site, that is, by the entropy increases. The hydrophobic interaction and enthalpy-entropy compensation at the binding site most likely cause the hydrophilic side-chains in this region to get rid of their original hydrogen-bonded water molecules, and promote formation of new hydrogen bonding and electrostatic attraction relationship among these hydrophilic residue-side chains at the binding site.

Mutation of some amino acid residues can reduce the hydrophobic surface areas of the SARS-CoV-2 RBD at the docking site and may significantly decrease the hydrophobic interaction between of SARS-CoV-2 S and ACE2, thereby greatly reducing the affinity between them. By analyzing the hydrophobic side-chains at the binding site of the complex of the SARS-CoV-2 RBD and ACE2, we tried to mutate the 6 amino acid residues to aspartame Acid in the RBD, see Fig. 8. For only 6 amino acid residues are mutated, the tertiary structure of the main chain of the mutated RBD may be the same as that of the original RBD. We simulated the molecular structure of the complex of the mutated RBD and ACE2 by using molecular dynamics NVT ensemble and NVERE relaxation algorithm (25). The simulation results show that the hydrophobic interaction areas of the mutant RBD in docking with ACE2 is greatly reduced by about 50%, which is similar to the size of hydrophobic interacting area of the SARS-CoV RBD bound to ACE2. Through such a mutation method, only a few amino acid residues mutation most likely can greatly reduce the affinity of the virus and the receptor, which may significantly reduce its infectiousness. This method may be used to design an attenuated virus that is very similar to origin coronavirus, but most likely retains its immunogenicity and triggers the immune response. By mutating several amino acid residues of SARS-CoV-2 RBD, the hydrophobic interaction of SARS-CoV-2 RBD and ACE2 can be significantly disrupted, thereby significantly reducing the binding efficiency of the virus to the host cell, which may help to slow down SARS-CoV-2 transmission from person to person.

Recently, molecular structures of a complex of human monoclonal antibody CB6 and SARS-CoV-2 RBD have been experimentally determined (26). By analyzing the distribution of the hydrophobic surface areas at the binding site of the complex, we found that the contact hydrophobic surface area of the RBD and CB6 is not big enough to exhibit the blocking effect and neutralizing capacity of the antibody to the virus. However, it is worth noting that the docking site of antibody CB6 is connected to a large hydrophobic area, as shown in Fig. 9A. This large hydrophobic region most likely contribute to hydrophobic interaction between the RBD and CB6 as an entropy-driven spontaneous process, thereby strengthening the binding of the CB6 and RBD. Molecular structures of another complex of human monoclonal antibody B38 and RBD of SARS-CoV-2 have also been experimentally determined (*31*). At the docking site, antibody B38 is also connected to a large hydrophobic area, as shown in Fig. 9B. Therefore, in evaluating hydrophobic interactions among proteins, hydrophobic surface areas connected with the docking site should be taken into consideration for the hydrophobic effect.

**Conclusion**

The high affinity between SARS-CoV-2 S and ACE2 most likely resourced from hydrophobic effect among the hydrophobic surface areas of the two protein at the binding site. The hydrophobic interaction and enthalpy-entropy compensation in the binding region between the S protein and ACE2 protein most likely cause the hydrophilic residues in this region to get rid of the hydrogen-bonded water molecules, and to promote hydrogen bonding and electrostatic attraction among these hydrophilic side-chains at the binding site. The hydrophobic portions of the hydrophilic side chains at the docking site participate in the hydrophobic interaction between SARS-CoV-2 S and ACE2. The affinity between SARS-CoV-2 RBD and ACE2 can be characterized by calculation the hydrophobic contact area between them at the docking site. This method shows that the hydrophobic interaction between the SARS-CoV-2 S and ACE2 protein is significantly greater than that between SARS-CoV S and ACE2. The degree of hydrophobic paring between SARS-CoV-2 RBD and ACE2 is very high. This explains why the affinity of SARS-CoV-2 RBD and ACE2 far exceeds that of SARS-CoV RBD and ACE2. Only several amino acid residues mutation may greatly reduce the affinity of SARS-CoV-2 and ACE2 receptor, which may significantly reduce its infectiousness. This method may be used to design an attenuated virus that is very similar to origin coronavirus, but still retains its immunogenicity and triggers the immune response. In evaluating hydrophobic interaction between virus and the receptor, hydrophobic surface areas connected with the binding sites should be taken into consideration that most likely play a role of increasing the hydrophobic effect in their docking.

## Materials and Methods

### Protein structures

In this study, many experimentally determined native structures of proteins are used to study the mechanism triggering docking of SARS-CoV-2 S and ACE2. All the three-dimensional (3D) structure data of protein molecules are resourced from the PDB database, including the experimentally determined the RBD of SARS-CoV-2 S, RBD of SARS-CoV S, ACE2, antibody CB6 and their complexes, et al. IDs of these proteins according to PDB database are marked in the Fig.1, Fig.3, Fig.4, Fig.5, Fig.6, Fig.7, Fig.8, and Fig.9. In order to show the distribution of hydrophobic areas on the surface of the SARS-CoV-2 RBD, of SARS-CoV RBD, ACE2, antibody and their complexes at the binding sites in these figures, we used the structural biology visualization software PyMOL to display the protein hydrophobic surface areas.

### MD simulations

The simulation of the hydration shells for the RBD of SARS-CoV-2 S and ACE2 was executed using NAMD simulator (*33*) with the CHARMM36 potential (*34, 35*) in an NVT ensemble at 300K for 5000000 time steps (2 fs per time step). Water molecules were built in these models 10Å away from the two protein structures. In the simulations, the hydration shells were gradually formed surrounding these structures. The shapes of hydration shells were achieved by showing water molecules within 3Å distance of the proteins' surfaces. The main chain structures of these model have no change during the hydration shells simulations. The van der Waals (VDW) interaction was truncated at 10 Å. We tried to mutate 6 amino acid residues of the RBD region of the complex of the spiked S protein of the COVID-19 virus and the RBD region of the docking complex of ACE2 protein. By using the molecular dynamics NVT ensemble (300K) and the NVERE relaxation algorithm (*36*), we simulated the molecular structure of the complex of docked spike protein and ACE2 protein. The simulation results shows that the main chain structure of the complex do not change due to mutation. The NVERE relaxation features the optimization of potential energy through long MD trajectories and large deformation, and it is capable of finding more stable equilibrium configurations than common optimization algorithms (*36*).

**Calculation of hydrophobic surface area of proteins involved in docking**

Affinity of RBD and ACE2 can be characterized by calculating the size of the hydrophobic contact area in the complex structures. The hydrophobic interaction regulating the docking of S protein and ACE2 mainly occurs at the docking sit. We used molecular 3D structure display software PyMOL to draw the hydrophobic surface areas which at least contacting another hydrophobic surface area at the docking site. Since these hydrophobic surface areas are very close to each other, we think these hydrophobic surfaces participate in the hydrophobic effect. We calculated the hydrophobic surface areas involved in the hydrophobic interaction between RBD and ACE2 in this study.

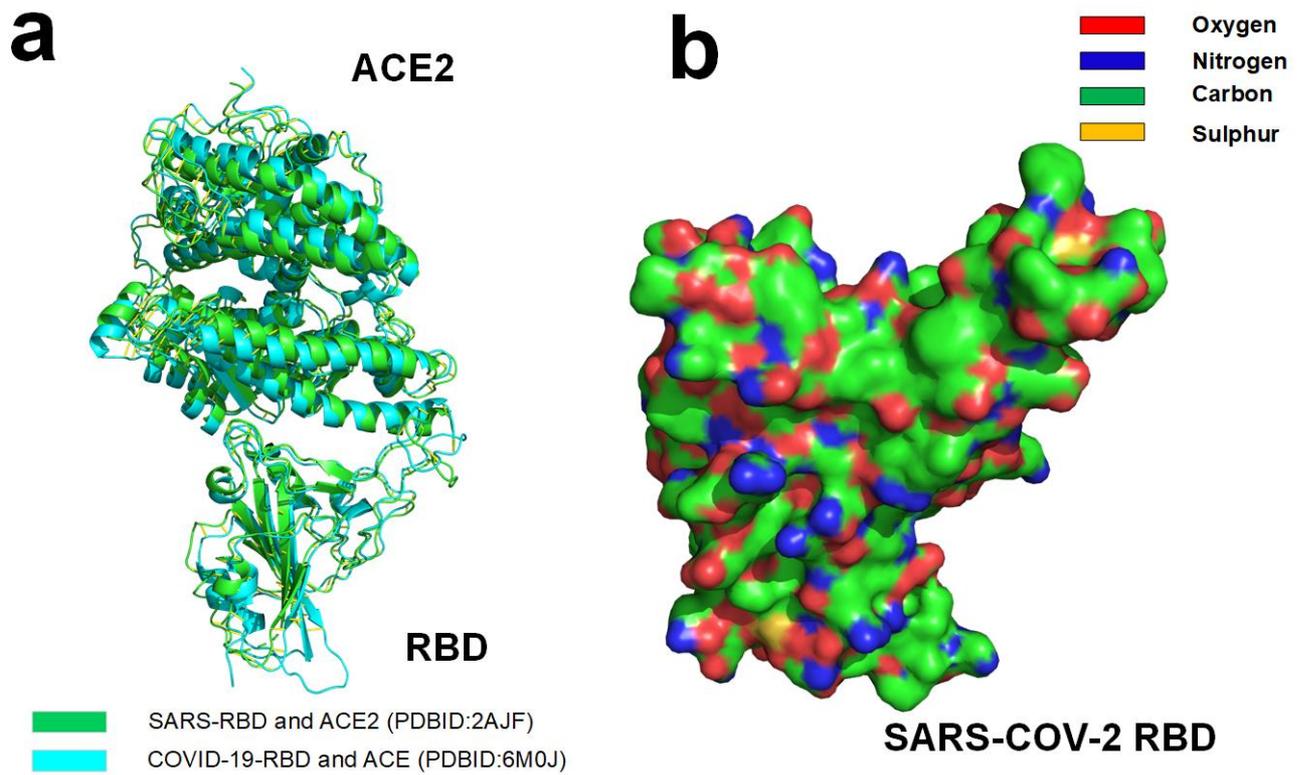

Fig.1 **a** Comparison of the complex of SARS-CoV-2 RBD bound to ACE2 and the complex of SARS-CoV RBD bound to ACE2(7) (11). **b** Molecular surface of SARS-CoV-2 RBD (hydrophobic surface areas is highlighted by green and yellow) (7)

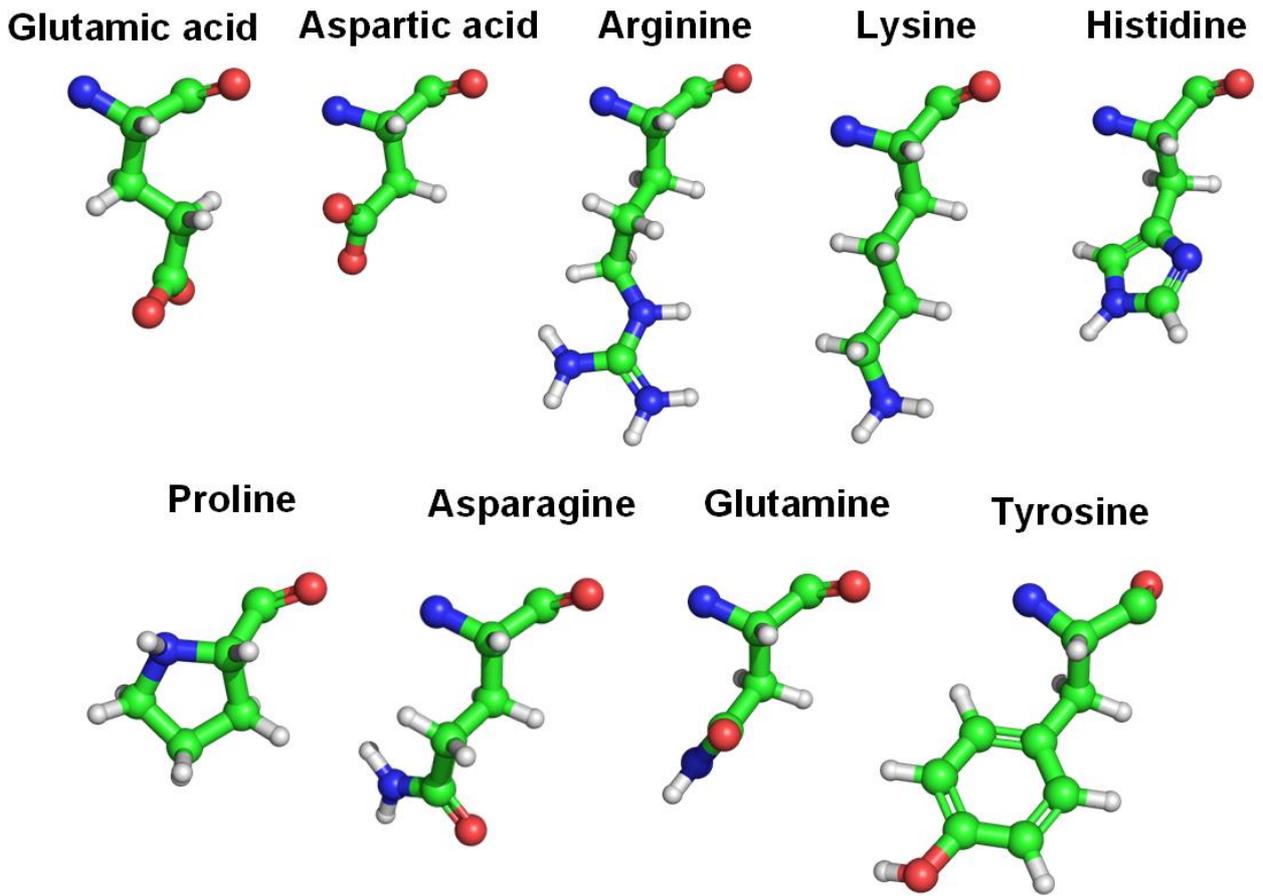

Fig.2 Hydrophobic portions of hydrophilic amino acid side-chains (hydrophobic portions are highlighted by green)

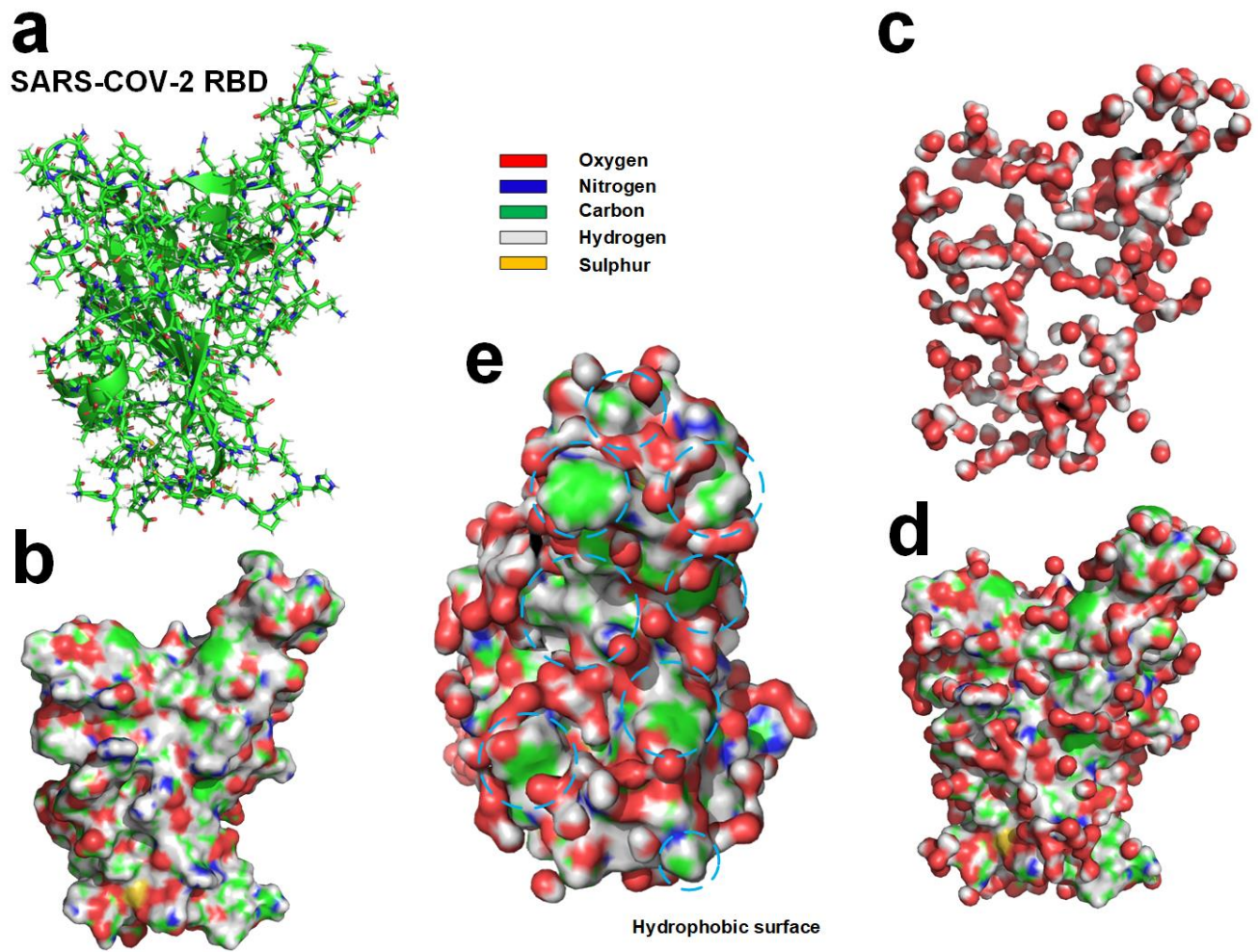

Fig.3 **a** Molecular structure of SARS-CoV-2 RBD (9) (PDBID: 6LZG) .**b** Molecular surface of SARS-CoV-2 RBD with supplementation of hydrogen atoms .**c** Hydrogen-bonded water molecules to the RBD. **d** The RBD and hydrogen-bonded water molecules. **e** Distribution of hydrogen-bonded water molecules at docking site of the RBD, the exposed hydrophobic surface is marked with dash lines.

*Corresponding author. E-mail address: linyang@hit.edu.cn (Lin Yang) [1]These authors contributed equally to this work.

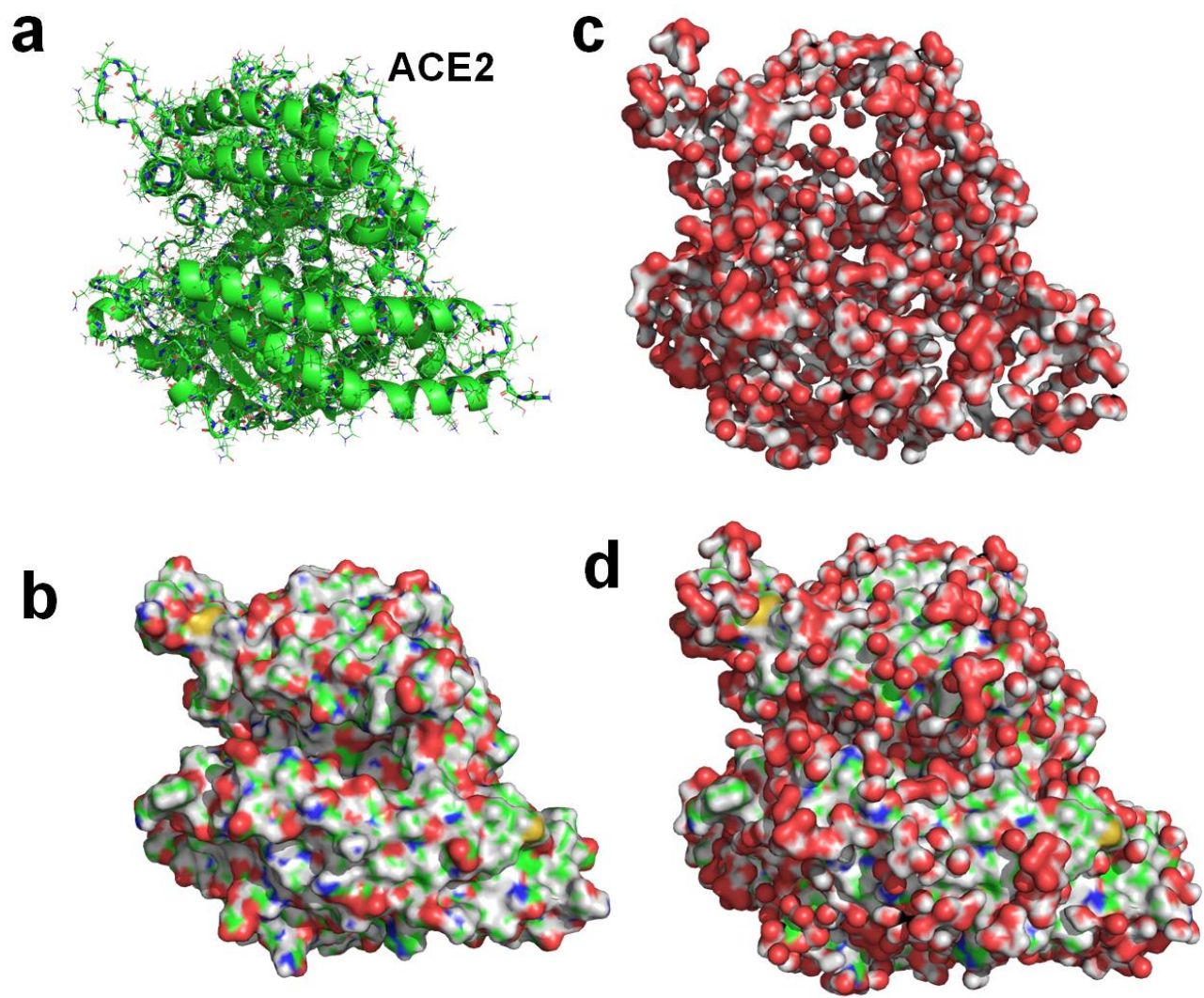

Fig.4 **a** Molecular structure of ACE2 (9) (PDBID: 6LZG). **b** Molecular surface of ACE2 protein with supplementation of hydrogen atoms. **c** Water molecules hydrogen-bonded to ACE2. **d** ACE2 and hydrogen-bonded water molecules.

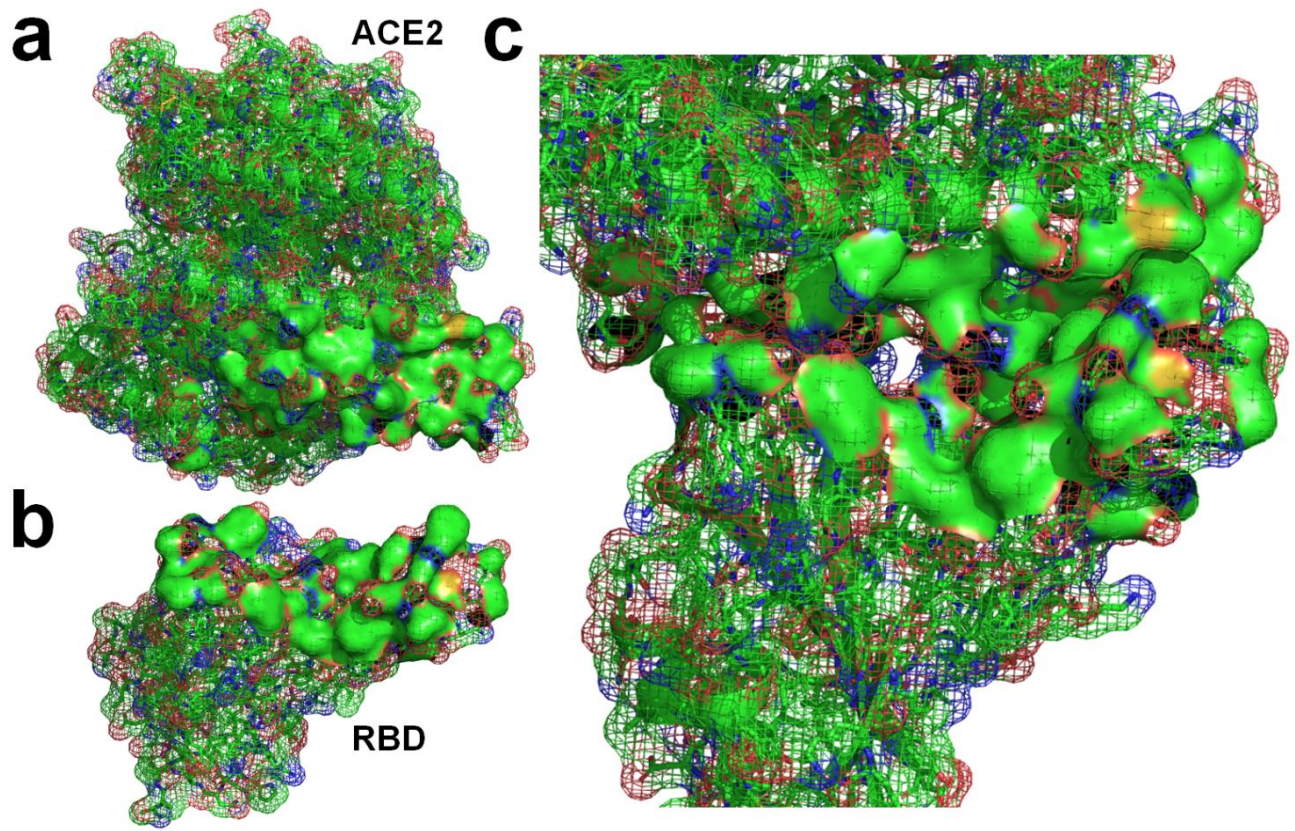

Fig.5 **a** Distribution of hydrophobic surface areas on the ACE2 involved in hydrophobic effect at the docking site (green surface areas) (PDBID: 6LZG). **b** Distribution of hydrophobic surface areas on the SARS-CoV-2 RBD involved in hydrophobic effect at the docking site (green surface areas), **c** The hydrophobic surface areas contacting in the complex of ACE2 and the RDB (green surface areas) (9)

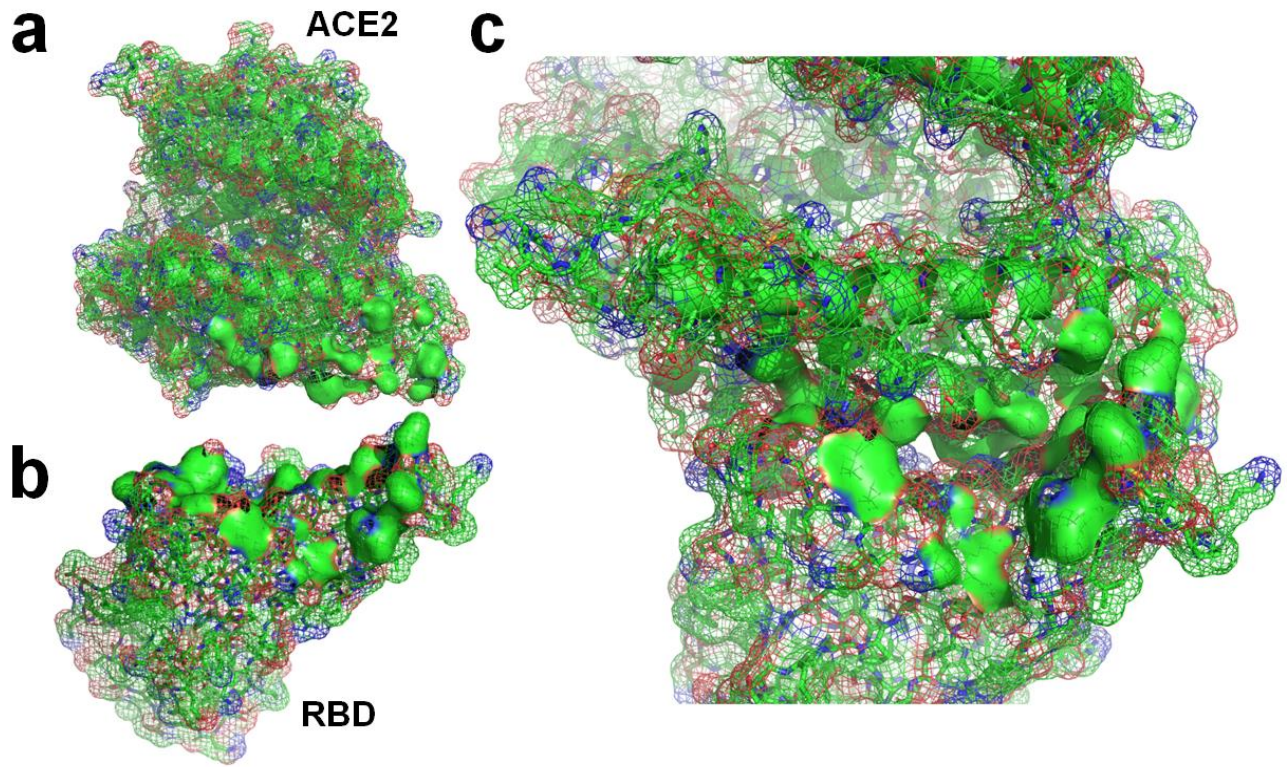

Fig.6 **a** Distribution of hydrophobic surface areas on the ACE2 involved in hydrophobic effect at the docking site (green surface areas) (PDBID: 2AJF), **b** Distribution of hydrophobic surface areas on the SARS-CoV RBD involved in hydrophobic effect at the docking site (green surface areas), **c** The hydrophobic surface areas contacting in the direction of docking in the complex of ACE2 and the RDB (green surface areas) (11)

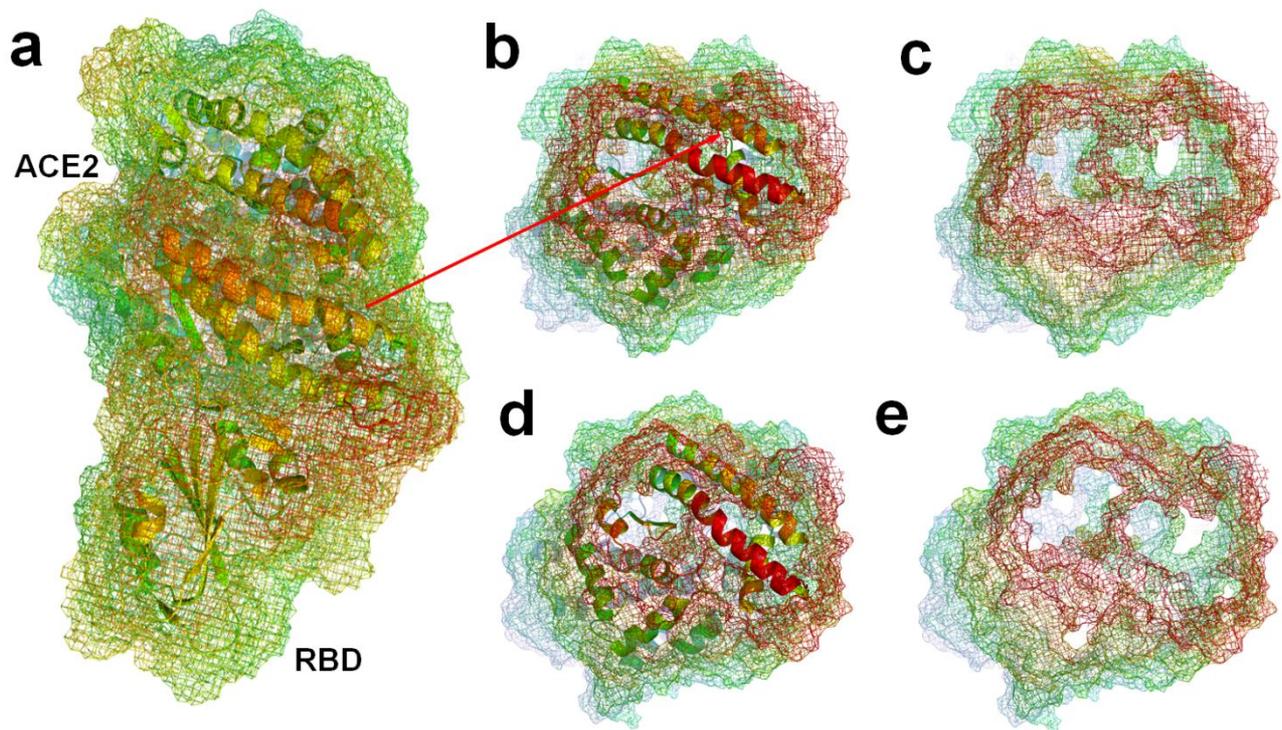

Fig.7 **a** Hydrated shell of SARS-CoV-2 RBD and ACE2, the grid represents the hydrated shell (PDBID: 6LZG) (9). **b ,c** cross-sectional views of the hydration shells at the docking site of SARS-CoV-2 RBD and ACE2 (PDBID: 6LZG), **e ,d** cross-sectional views of the hydration shells at the docking site of SARS-CoV RBD and ACE2 (PDBID: 2AJF) (11) .

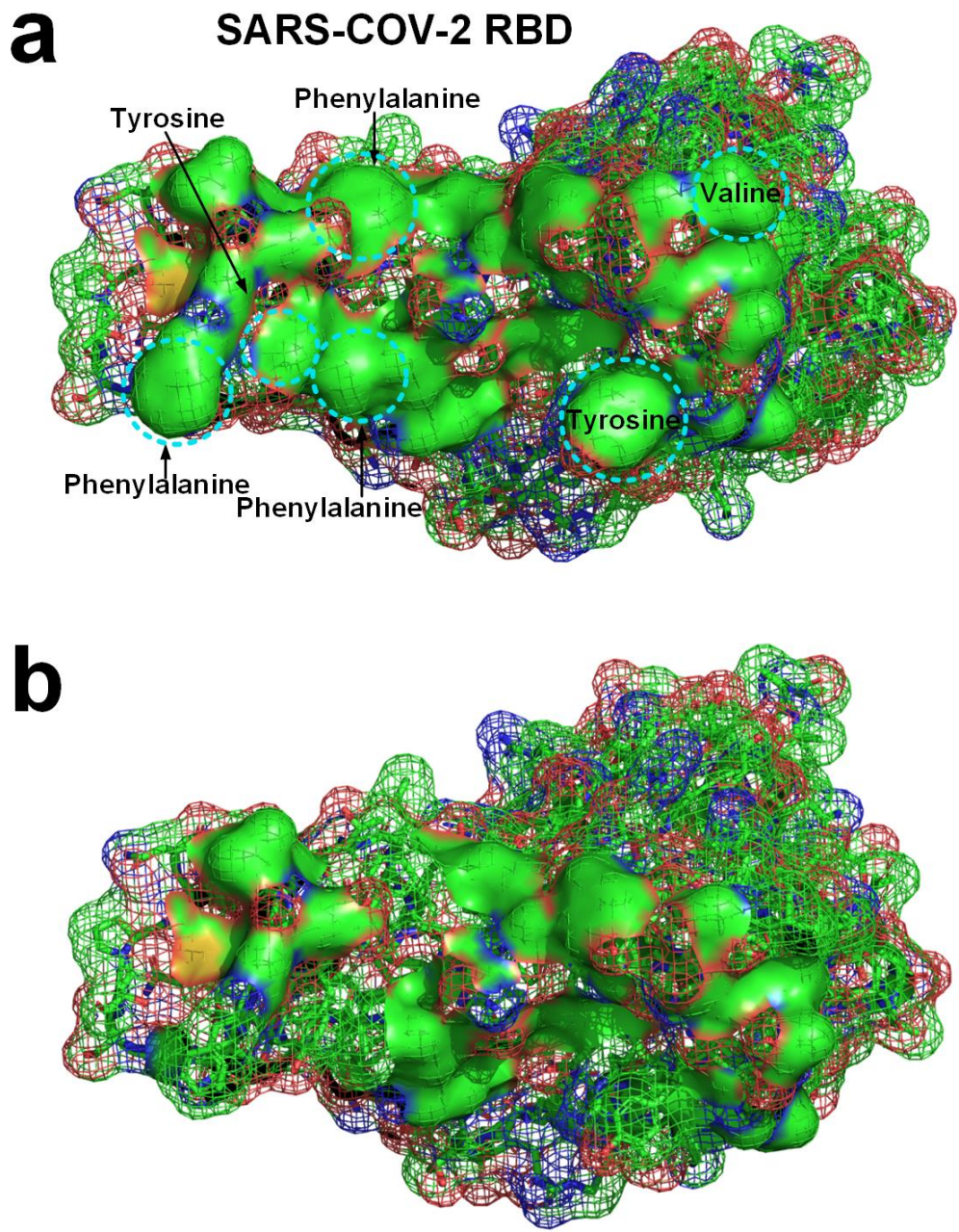

Fig.8 **a** Six hydrophobic residues in SARS-CoV-2 RBD at the docking site (PDBID: 6LZG) (9). **b** The distribution of hydrophobic surface areas of the RBD after mutating the 6 hydrophobic residues to aspartic Acid at the docking site.

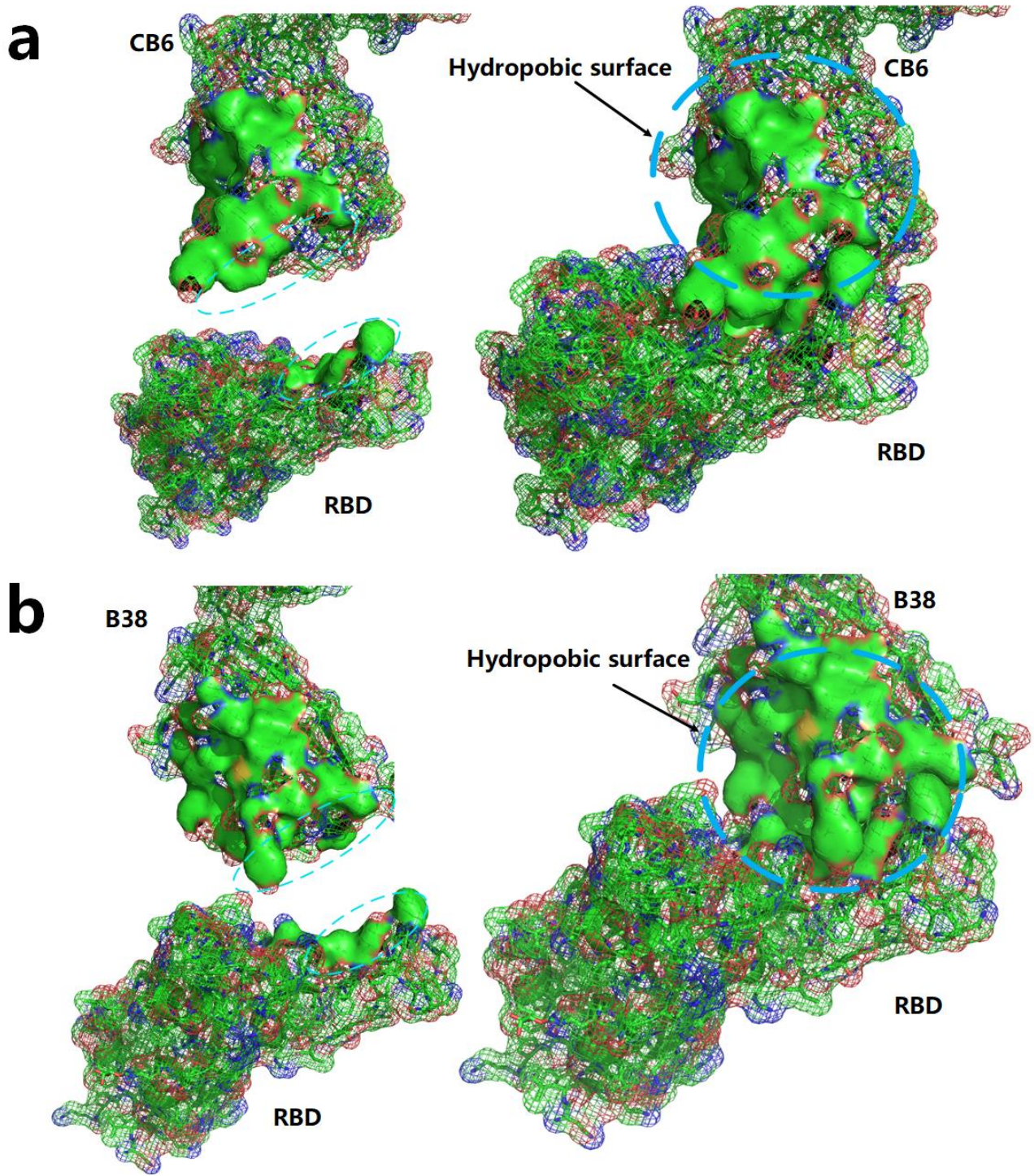

Fig.9 **a** The hydrophobic surface areas of the human monoclonal antibody CB6 and SARS-CoV-2 RBD at the binding site (green surface areas) (PDBID: 7C01 ) (37). **b** The hydrophobic surface areas of the human monoclonal antibody B38 and SARS-CoV-2 RBD at the binding site (green surface areas) (PDBID: 7C01 ) (*31*).

**Acknowledgement**

The authors acknowledge financial support from the Fundamental Research Funds for the Central Universities of China. Lin Yang is indebted to Daniel Wagner from the Weizmann Institute of Science and Liyong Tong from the University of Sydney for their support and guidance. Lin Yang is grateful for his research experience in the Weizmann Institute of Science for inspiration. The authors acknowledge the financial support from the National Natural Science Foundation of China (Grant 21601054), Shenzhen Science and Technology Program (Grant No. KQTD2016112814303055), Science Foundation of the National Key Laboratory of Science and Technology on Advanced Composites in Special Environments and the University Nursing Program for Young Scholars with Creative Talents in Heilongjiang Province of China (Grants UNPYSCT-2017126). The authors thank Joan Rosenthal for editing this paper.

**Additional Information**

The authors declare no competing financial interests.